\documentclass[runningheads]{llncs}
\usepackage[colorlinks=true,linkcolor=blue,urlcolor=blue,citecolor=blue]{hyperref}
\usepackage{graphicx}
 \usepackage{amsmath}
 \usepackage{booktabs}
 \usepackage{multirow}
 \usepackage{amsfonts}
 \usepackage{float}
 \usepackage{caption}
 \usepackage{algorithm}
\usepackage{algorithmicx}
\usepackage{algpseudocode} 
\usepackage{subfigure}
\usepackage{cleveref}
\usepackage{xcolor}
\usepackage[misc,geometry]{ifsym}

\begin{document}
\title{CoLa-Diff: Conditional Latent Diffusion Model for Multi-Modal MRI Synthesis\thanks{L. Jiang and Y. Mao contribute equally in this work}}
\titlerunning{CoLa-Diff for Multi-Modal MRI Synthesis}
\author{Lan Jiang\inst{1} \and
Ye Mao\inst{2} \and
Xi Chen\inst{3} \and
Xiangfeng Wang\inst{4}\and
Chao Li\inst{1,2,5}\textsuperscript{\Letter}}

\authorrunning{L. Jiang et al.}
%
\institute{School of Science and Engineering, University of Dundee \and
Department of Clinical Neurosciences, University of Cambridge
\and
Department of Computer Science, University of Bath 
\and
School of Computer Science and Technology, East China Normal University
\and 
School of Medicine, University of Dundee
\\
\email{cl647@cam.ac.uk}
}
%
%
\maketitle              

\begin{abstract}
 MRI synthesis promises to mitigate the challenge of missing MRI modality in clinical practice. Diffusion model has emerged as an effective technique for image synthesis by modelling complex and variable data distributions. However, most diffusion-based MRI synthesis models are using a single modality. As they operate in the original image domain, they are memory-intensive and less feasible for multi-modal synthesis. Moreover, they often fail to preserve the anatomical structure in MRI. Further, balancing the multiple conditions from multi-modal MRI inputs is crucial for multi-modal synthesis. Here, we propose the first diffusion-based multi-modality MRI synthesis model, namely Conditioned Latent Diffusion Model (CoLa-Diff). To reduce memory consumption, we design CoLa-Diff to operate in the latent space. We propose a novel network architecture, e.g., similar cooperative filtering, to solve the possible compression and noise in latent space. To better maintain the anatomical structure, brain region masks are introduced as the priors of density distributions to guide diffusion process. We further present auto-weight adaptation to employ multi-modal information effectively. Our experiments demonstrate that CoLa-Diff outperforms other state-of-the-art MRI synthesis methods, promising to serve as an effective tool for multi-modal MRI synthesis.

\keywords{Multi-modal MRI \and Medical image synthesis \and Latent space \and Diffusion models \and Structural guidance }
\end{abstract}
\section{Introduction}
Magnetic resonance imaging (MRI) is critical to the diagnosis, treatment, and follow-up of brain tumour patients \cite{MRI_theory_practice}. Multiple MRI modalities offer complementary information for characterizing brain tumours and enhancing patient management \cite{multimodal_important,multi_modal_learning}. However, acquiring multi-modality MRI is time-consuming, expensive and sometimes infeasible in specific modalities, e.g., due to the hazard of contrast agent \cite{contrast_agents}. Trans-modal MRI synthesis can establish the mapping from the known domain of available MRI modalities to the target domain of missing modalities, promising to generate missing MRI modalities effectively.
\\ ${\rm{     }}$ ${\rm{     }}$ ${\rm{     }}$ ${\rm{     }}$
The synthetic methods leveraging multi-modal MRI, i.e., many-to-one translation, have outperformed single-modality models generating a missing modality from another single available modality, i.e., one-to-one translation \cite{zhou2020hi,MM-GAN}. Traditional multi-modal methods \cite{traditional_1,traditional_2}, e.g., sparse encoding-based, patch-based and atlas-based methods, highly rely on the alignment accuracy of source and target domains and are poorly scalable. Recent generative adversarial networks (GANs) and variants, e.g., MM-GAN \cite{MM-GAN}, DiamondGAN \cite{diamondgan} and ProvoGAN \cite{ProvoGan}, have been successful based on multi-modal MRI, further improved by introducing multi-modal coding \cite{gate_mergence}, enhanced architecture \cite{ResViT}, and novel learning strategies \cite{mousegan++}. 
\\ ${\rm{     }}$ ${\rm{     }}$ ${\rm{     }}$ ${\rm{     }}$
Despite the success, GAN-based models are challenged by the limited capability of adversarial learning in modelling complex multi-modal data distributions \cite{mode_collapse} Recent studies have demonstrated that GANs' performance can be limited to processing and generating data with less variability \cite{gan_cannot}. In addition, GANs' hyperparameters and regularization terms typically require fine-tuning, which otherwise often results in gradient vanish and mode collapse \cite{optimization_gan}. 
\\ ${\rm{     }}$ ${\rm{     }}$ ${\rm{     }}$ ${\rm{     }}$
Diffusion model (DM) has achieved state-of-the-art performance in synthesizing natural images, promising to improve MRI synthesis models. It shows superiority in model training \cite{DMbeatGan}, producing complex and diverse images \cite{improvedDDPM,DDPM}, while reducing risk of modality collapse \cite{no_modecollapse_ddpm}.For instance, Lyu et al \cite{lyuCTMRI} used diffusion and score-marching models to quantify model uncertainty from Monte-Carlo sampling and average the output using different sampling methods for CT-to-MRI generation; Özbey et al \cite{syndiff} leveraged adversarial training to increase the step size of the inverse diffusion process and further designed a cycle-consistent architecture for unpaired MRI translation. 
\\ ${\rm{     }}$ ${\rm{     }}$ ${\rm{     }}$ ${\rm{     }}$ 
However, current DM-based methods focus on one-to-one MRI translation, promising to be improved by many-to-one methods, which requires dedicated design to balance the multiple conditions introduced by multi-modal MRI. Moreover, as DMs operate in original image domain, all Markov states are kept in memory \cite{DDPM}, resulting in excessive memory burden and reduced feasibility of many-to-one translation. Further, diffusion denoising processes tend to change the original distribution structure of the target image due to noise randomness \cite{lyuCTMRI}, rending DMs often ignore the consistency of anatomical structures embedded in medical images, leading to clinically less relevant results. Lastly, DMs are known for their slow speed of diffusion sampling \cite{DDPM,improvedDDPM,fastsample}, challenging its wide clinical application. 
  \\ ${\rm{     }}$ ${\rm{     }}$ ${\rm{     }}$ ${\rm{     }}$
We propose a DM-based multi-modal MRI synthesis model, CoLa-Diff, which facilitates many-to-one MRI translation in latent space, and preserve anatomical structure with accelerated sampling. Our main contributions include: 
\begin{itemize}
\item We present a denoising diffusion probabilistic model based on multi-modal MRI. As far as we know, this is the first DM-based many-to-one MRI synthesis model. 
\item We design a bespoke architecture to facilitate diffusion operations in the latent space, e.g., similar cooperative filtering, to reduce the risks of excessive information compression and high-dimensional noise in the latent space.
\item  We introduce structural guidance of brain regions in each step of the diffusion process, preserving anatomical structure and enhancing synthesis quality. 
\item  We propose an approach for adapting condition weights automatically to balance multiple conditions and maximise the chance of leveraging relevant multi-modal information.
\end{itemize}

\section{Multi-conditioned Latent Diffusion Model}
\begin{figure}[t]
\includegraphics[width=\textwidth]{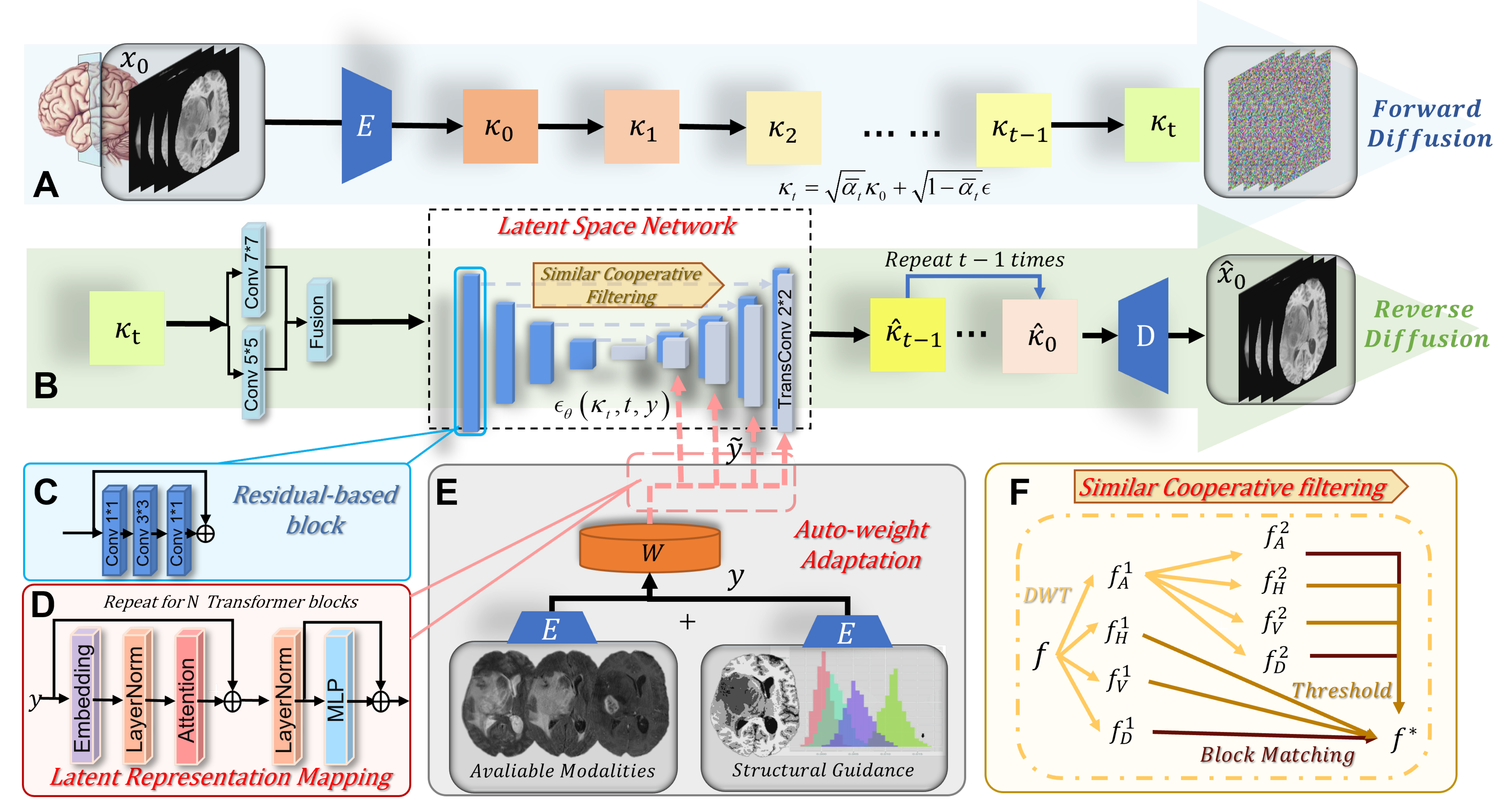}
\caption{Schematic diagram of CoLa-Diff. During the forward diffusion, Original images $x_0$ are compressed using encoder $E$ to get $\kappa_0$, and after $t$ steps of adding noise, the images turn into $\kappa_t$. During the reverse diffusion, the latent space network $\epsilon_\theta\left(\kappa_t, t, y\right)$ predicts the added noise, and other available modalities and anatomical masks as structural guidance are encoded to $y$, then processed by the auto-weight adaptation block $W$ and embedded into the latent space network. Sampling from the distribution learned from the network gives $\hat{\kappa}_{0}$, then $\hat{\kappa}_{0}$ are decoded by $D$ to obtain synthesized images.} \label{Fig.1}
\end{figure}

 Fig.\ref{Fig.1} illustrates the model design. As a latent diffusion model, CoLa-diff integrates multi-condition $b$ from available MRI contrasts in a compact and low-dimensional latent space to guide the generation of missing modality $x\in\mathbb{R}^{H \times W \times 1}$. Precisely, $b$ constitutes available contrasts and anatomical structure masks generated from the available contrasts. 
Similar to \cite{DDPM}, CoLa-Diff invovles a forward and a reverse diffusion process. During forward diffusion, $x_0$ is encoded by $E$ to produce $\kappa_0$, then subjected to $T$ diffusion steps to gradually add noise $\epsilon$ and generate a sequence of intermediate representations: $\{\kappa_0,\ldots,\kappa_{T}\}$. The $t$-th intermediate representation is denoted as $\kappa_{t}$, expressed as:
\begin{equation}
\kappa_t=\sqrt{\bar{\alpha}_t} \kappa_0+\sqrt{1-\bar{\alpha}_t} \epsilon, \quad \text { with } \epsilon \sim \mathcal{N}(0, \mathbf{I})
\end{equation}
where ${{\bar \alpha }_t} = \prod\nolimits_{i = 1}^t {{\alpha _i}}$, ${{\alpha }_i}$ denotes hyper-parameters related to variance. 

The reverse diffusion is modelled by a latent space network with parameters $\theta$, taking intermediate perturbed feature maps $\kappa_t$ and $y$ (compressed $b$) as input to predict a noise level $\epsilon_\theta\left(\kappa_t, t, y\right)$ for recovering feature maps $\hat\kappa_{t-1}$ from previous steps,
\begin{equation}
    {\hat\kappa _{t - 1}} = \sqrt {{{\bar \alpha }_{t - 1}}} (\frac{{{\kappa _t} - \sqrt {1 - {{\bar \alpha }_t}}  \cdot \epsilon{_\theta }\left( {{\kappa _t},t,y} \right)}}{{\sqrt {{{\bar \alpha }_t}} }}) + \sqrt {1 - {{\bar \alpha }_{t - 1}}}  \cdot \epsilon{_\theta }\left( {{\kappa _t},t,y} \right)
    \label{Eq.2}
\end{equation}

To enable effective learning of the underlying distribution of $\kappa_0$, the noise level needs to be accurately estimated. To achieve this, the network employs similar cooperative filtering and auto-weight adaptation strategies. $\hat\kappa_0$ is recovered by repeating Eq. \ref{Eq.2} process for $t$ times, and decoding the final feature map to generate synthesis images $\hat{x_0}$.

\subsection{Latent Space Network} 
We map multi-condition to the latent space network for guiding noise prediction at each step $t$. The mapping is implemented by $N$ transformer blocks (Fig.\ref{Fig.1} (D)), including global self-attentive layers, layer-normalization and position-wise MLP.
The network $\epsilon_\theta\left(\kappa_t, t, y\right)$ is trained to predict the noise added at each step using
\begin{equation}
\mathcal{L}_{\mathrm{E}}:=\mathbb{E}_{{E}(x), y, \epsilon \sim \mathcal{N}(0,1), t}\left[\left\|\epsilon-\epsilon_\theta\left(\kappa_t, t, y\right)\right\|_2^2\right]
\end{equation}

To mitigate the excessive information losses that latent spaces are prone to, we replace the simple convolution operation with a residual-based block (three sequential convolutions with kernels $1*1$, $3*3$, $1*1$ and residual joins \cite{resnet}), and enlarge the receptive field by fusion ($5*5$ and $7*7$ convolutions followed by AFF \cite{AFF}) in the down-sampling section. Moreover, to reduce high-dimensional noise generated in the latent space, which can significantly corrupt the quality of multi-modal generation.  we design a similar cooperative filtering detailed below.

\subsubsection{Similar Cooperative filtering} The approach has been devised to filter the downsampled features, with each filtered feature connected to its respective upsampling component (shown in Fig.\ref{Fig.1} (F)). Given $f$, which is the downsampled feature of $\kappa_t$, suppose the 2D discrete wavelet transform $\phi$ \cite{DWT} decomposes the features into low frequency component $f_A^{(i)}$  and high frequency components $f_H^{(i)}$, $f_V^{(i)}$, $f_D^{(i)}$, keep decompose $f_A^{(i)}$, where $i$ is the number of wavelet transform layers. We group the components and further filter by similar block matching $\delta$ \cite{block_match} or thresholding $\gamma$, use the inverse wavelet transform ${\phi ^{ - 1}(\cdot)}$ to reconstruct the denoising results, given $f^*$.
\begin{equation}
 {f^*} = {\phi ^{ - 1}}(\delta (f_A^{(i)}),\delta (\sum\limits_{j = 1}^i {f_D^{(i)}} ),\gamma (\sum\limits_{j = 1}^i {f_H^{(i)}} ),\gamma (\sum\limits_{j = 1}^i {f_V^{(i)}} ))  
\end{equation}

\subsection{Structural Guidance}
Unlike natural images, medical images encompass rich anatomical information. Therefore, preserving anatomical structure is crucial for MRI generation. However, DMs often corrupt anatomical structure, and this limitation could be due to the learning and sampling processes of DMs that highly rely on the probability density function \cite{DDPM}, while brain structures by nature are overlapping in MRI density distribution and even more complicated by pathological changes. 


Previous studies show that introducing geometric priors can significantly improve the robustness of medical image generation. \cite{structural_guide_2,structural_edge}. Therefore, we hypothesize that incorporating structural prior could enhance the generation quality with preserved anatomy. Specifically, we exploit FSL-FAST \cite{FAST} tool to segment four types of brain tissue: white matter, grey matter, cerebrospinal fluid, and tumour. The generated tissue masks and inherent density distributions (Fig.\ref{Fig.1} (E)) are then used as a condition $y_i$ to guide the reverse diffusion. 

The combined loss function for our multi-conditioned latent diffusion is defined as
\begin{equation}
    \mathcal{L}_{\mathrm{MCL}}:= \mathcal{L}_{\mathrm{E}}+\mathcal{L}_{\mathrm{KL}}
\end{equation}
where $\mathrm{KL}$ is the KL divergence loss to measure similarity between real $q$ and predicted $p_\theta$ distributions of encoded images.

\begin{equation}
\mathcal{L}_{\mathrm{KL}}:=\sum_{j=1}^{T-1} D_{K L}\left(q\left(\kappa_{j-1} \mid \kappa_j, \kappa_0\right) \| p_\theta\left(\kappa_{j-1} \mid \kappa_j\right)\right)
\end{equation}
where $D_{\mathrm{KL}}$ is the KL divergence function. 


\subsection{Auto-weight adaptation} It is critical to balance multiple conditions, maximizing relevant information and minimising redundant information. For encoded conditions $y \in {\mathbb{R}^{h \times w \times c}}$, $c$ is the number of condition channels. Set the value after auto-weight adaptation to ${\tilde y}$, the operation of this module is expressed as (shown in Fig.\ref{Fig.1} (E))

\begin{equation}
\tilde y = F(y|\mu ,\nu ,o), \quad \text { with } \mu ,\nu ,o \in \mathbb{R}{^c}
\end{equation}

The embedding outputs are adjusted by embedding weight $\mu$. The auto-activation is governed by the learnable weight $\nu$ and bias $o$. $y_c$ indicates each channel of $y$, where ${y_c} = {[y_c^{m,n}]_{h \times w}} \in {R^{h \times w}}$, $y_c^{m,n}$ is the eigenvalue at position $(m,n)$ in channel $c$. We use large receptive fields and contextual embedding to avoid local ambiguities, providing embedding weight $\mu=[\mu_1, \mu_2 ..., \mu_c]$. The operation $G_c$ is defined as:

\begin{equation}
G_c=\mu_c\left\|y_c\right\|_2=\mu_c\left\{\left[\sum_{m=1}^h \sum_{n=1}^w\left(y_c^{m, n}\right)^2\right]+\varpi \right\}^{\frac{1}{2}}
\end{equation}
where $\varpi$ is a small constant added to the equation to avoid the issue of derivation at the zero point. The normalization method can establish stable competition between channels, $\mathbf{G}=\{G_c\}_{c=1}^{S}$. We use $L_2$ normalization for cross-channel operations:
\begin{equation}
\hat{G}_c=\frac{\sqrt{S} G_c}{\|\mathbf{G}\|_2}=\frac{\sqrt{S} G_c}{\left[\left(\sum_{c=1}^S G_c^2\right)+\varpi\right]^{\frac{1}{2}}}
\end{equation}
where $S$ denotes the scale. We use an activation mechanism for updating each channel to facilitate the maximum utilization of each condition during diffusion model training, and further enhance the synthesis performance. Given the learnable weight $\mathbf{\nu}=[\nu_1,\nu_2,..., \nu_c]$ and bias $\mathbf{o}=[o_1,o_2,..., o_c]$ we compute
\begin{equation}
{{\tilde y}_c} = {y_c}[1 + S({\nu _c}{{\hat G}_c} + {o_c})]
\end{equation}
which gives new representations ${\tilde y}_c$ of each compressed conditions after the automatic weighting. $S(\cdot)$ denotes the Sigmoid activation function.



\section{Experiments and Results} 

\subsection{Comparisons with State-of-the-Art Methods}
\subsubsection{Datasets and Baselines}We evaluated  CoLa-Diff on two multi-contrast brain MRI datasets:  BRATS 2018 and IXI datasets. The BRATS 2018  contains MRI scans from 285 glioma patients. Each includes four modalities: T1, T2, T1ce, and FLAIR. We split them into (190:40:55 for training/validation/testing. For each subject, we automatically selected axial cross-sections based on the perceptible effective area of the slices, and then cropped the selected slices to a size of $224\times224$. The IXI\footnote{\url{https://brain-development.org/ixi-dataset/}} dataset consists of 200 multi-contrast MRIs from healthy brains, plit them into (140:25:35) for training/validation/testing. For preprocessing, we registered T2- and PD-weighted images to T1-weighted images using FSL-FLIRT \cite{FLIRT}, and other preprocessing are identical to the BRATS 2018.

We compared CoLa-Diff with four state-of-the-art multi-modal MRI synthesis methods: MM-GAN \cite{MM-GAN}, Hi-Net \cite{zhou2020hi}, ProvoGan\cite{ProvoGan} and LDM\cite{LDM}.


\subsubsection{Implementation Details} The hyperparameters of CoLa-Diff are defined as follows: diffusion steps to 1000; noise schedule to linear; attention resolutions to $32,16,8$; batch size to 8, learning rate to $9.6e-5$. The noise variances were in the range of ${\beta _1} = {10^{ - 4}}$ and ${\beta _T} = 0.02$.
An exponential moving average (EMA) over model parameters with a rate of $0.9999$ was employed. The model is trained on $2$ NVIDIA RTX A5000, 24 GB with Adam optimizer on PyTorch. An acceleration method \cite{fastsample} based on knowledge distillation was applied for fast sampling.
\begin{table}[t]
  \centering
  \caption{Performance in BRATS (top) and IXI (bottom). PSNR (dB) and SSIM (\%) are listed as mean±std in the test set. \textbf{Boldface} marks the top models.}
  \resizebox{\linewidth}{!}{
    \begin{tabular}{lllllllll}
    \toprule
    \multicolumn{1}{c}{\multirow{3}[4]{*}{Model (BRATS 2018)}} & \multicolumn{2}{c}{T2+T1ce+FLAIR} & \multicolumn{2}{c}{T1+T1ce+FLAIR } & \multicolumn{2}{c}{T2+T1+FLAIR } & \multicolumn{2}{c}{T2+T1ce+T1 } \\
          & \multicolumn{2}{c}{→T1} & \multicolumn{2}{c}{→T2} & \multicolumn{2}{c}{ →T1ce} & \multicolumn{2}{c}{→FLAIR} \\
\cmidrule{2-9}          & PSNR  & SSIM\% & PSNR  & SSIM\% & PSNR  & SSIM\% & PSNR  & SSIM\% \\
    \midrule
    MM-GAN & 25.78±2.16 & 90.67±1.45 & 26.11±1.62 & 90.58±1.39 & 26.30±1.91 & 91.22±2.08 & 24.09±2.14 & 88.32±1.98 \\
    Hi-Net & 27.42±2.58 & 93.46±1.75 & 25.64±2.01 & 92.59±1.42 & 27.02±1.26 & 93.35±1.34 & 25.87±2.82 & 91.22±2.13 \\
    ProvoGAN & 27.79±4.42 & 93.51±3.16 & 26.72±2.87 & 92.98±3.91 & 29.26±2.50 & 93.96±2.34 & 25.64±2.77 & 90.42±3.13 \\
    LDM   & 24.55±2.62 & 88.34±2.51 & 24.79±2.67 & 88.47±2.60 & 25.61±2.48 & 89.18±2.55 & 23.12±3.16 & 86.90±3.24 \\
    CoLa-Diff (Ours) & \textbf{28.26±3.13} & \textbf{93.65±3.02} & \textbf{28.33±2.27} & \textbf{93.80±2.75} & \textbf{29.35±2.40} & \textbf{94.18±2.46} & \textbf{26.68±2.74} & \textbf{91.89±3.11} \\
    \bottomrule
    \end{tabular}
    }
\resizebox{\linewidth}{!}{
    \begin{tabular}{lllllllll}
    \toprule
    \multicolumn{1}{c}{\multirow{2}[4]{*}{Model (IXI)}} & \multicolumn{2}{c}{T1+T2 →PD} & \multicolumn{2}{c}{T2+PD →T1} & \multicolumn{2}{c}{T1+PD →T2} \\
\cmidrule{2-7}          & PSNR  & SSIM\% & PSNR  & SSIM\% & PSNR  & SSIM\% \\
    \midrule
    MM-GAN & 30.61±1.64 & 95.42±1.90 & 27.32±1.70 & 92.35±1.58 & 30.87±1.75 & 94.68±1.42 \\
    Hi-Net & 31.79±2.26 & 96.51±2.03 & 28.89±1.43 & 93.78±1.31 & 32.58±1.85 & 96.54±1.74 \\
    ProvoGAN & 29.93±3.11 & 94.62±2.40 & 24.21±2.63 & 90.46±3.58 & 29.19±3.04 & 94.08±3.87 \\
    LDM   & 27.36±2.48 & 91.52±2.39 & 24.19±2.51 & 88.75±2.47 & 27.04±2.31 & 91.23±2.24 \\
    CoLa-Diff (Ours) & \textbf{32.24±2.95} & \textbf{96.95±2.26} & \textbf{30.20±2.38} & \textbf{94.49±2.15} & \textbf{32.86±2.83} & \textbf{96.57±2.27} \\
    \bottomrule
    \end{tabular}
    }
  \label{Table1}%
\end{table}%

\subsubsection{Quantitative Results} We performed synthesis experiments for all modalities, with each modality selected as the target modality while remaining modalities and the generated region masks as conditions. Seven cases were tested in two datasets (Table \ref{Table1}). The results show that CoLa-Diff outperforms other models by up to 6.01 dB on PSNR and 5.74\% on SSIM. Even when compared to the best of other models in each task, CoLa-Diff is a maximum of 0.81 dB higher in PSNR and 0.82\% higher in SSIM.

\begin{figure}[H]
\centering
\includegraphics[width=0.9\textwidth,height=0.58\textwidth]{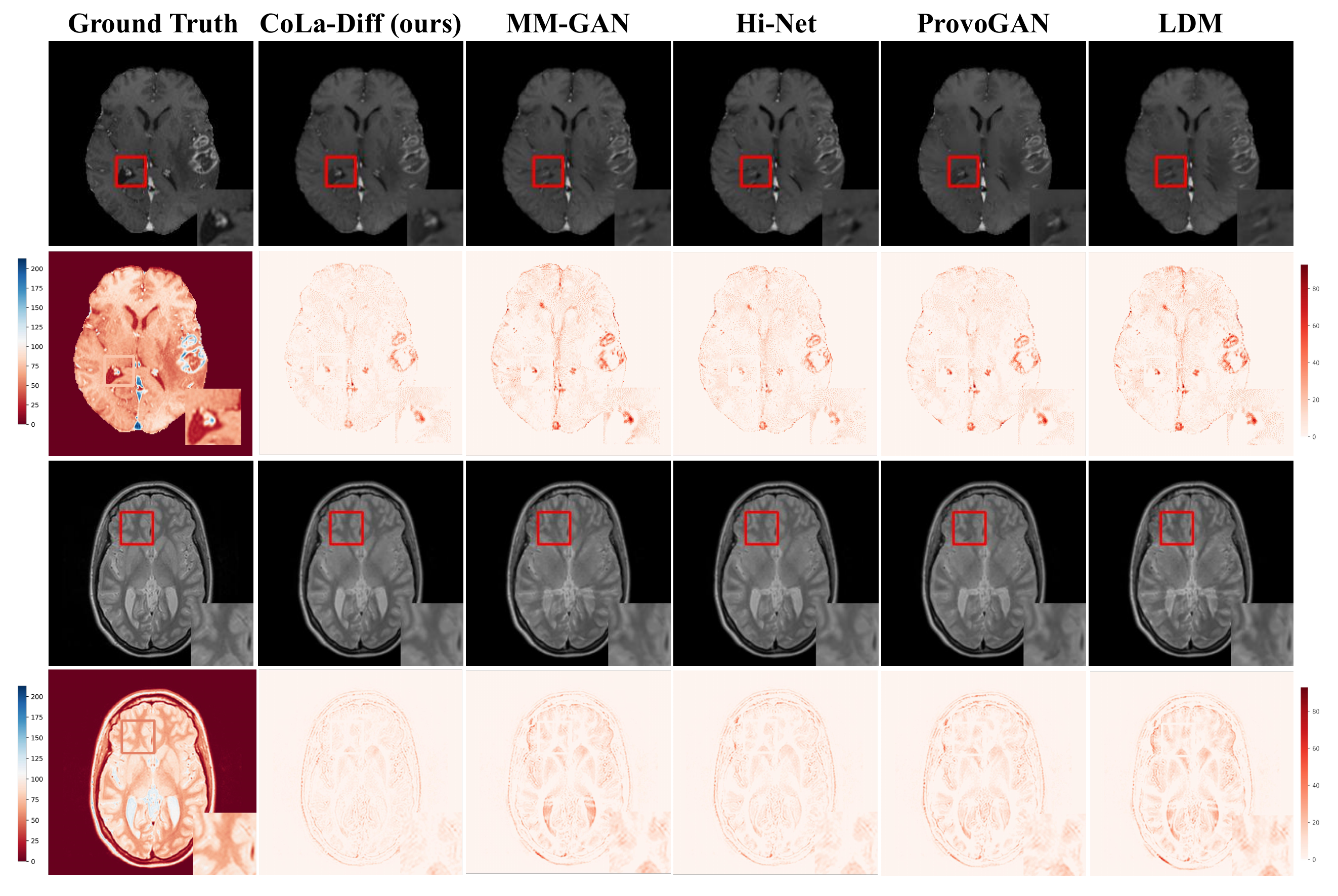}\caption{Visualization of synthesized images, detail enlargements (row 1 and 3) and corresponding error maps (row 2 and 4). } \label{fig2}
\end{figure}

\subsubsection{Qualitative results} The first two rows in Fig.\ref{fig2} illustrate the T1ce synthesis results on the BRATS dataset. The last two rows show the PD synthesis results on the IXI dataset. From the generated images, we observe that CoLa-Diff is most comparable to ground truth, with fewer errors shown in the heat maps. CoLa-Diff performs particularly better in generating complex brain sulcus and tumour boundaries. Further, CoLa-Diff can better maintain the anatomical structure of the original image over other comparison models.

\subsection{Ablation Study and Multi-modal Exploitation Capabilities}
We verified the effectiveness of each component in CoLa-Diff by removing them individually. We experimented on BRATS T1+T1ce+FLAIR→T2 task with four absence scenarios (Table \ref{ablation} top). Our results show that each component contributes to the performance improvement, with Auto-weight adaptation bringing a PSNR increase of 1.9450dB and SSIM of 4.0808\%.


\begin{table}
  \centering
  \caption{Ablation of four individual components (First four lines) and Multi-modal information utilisation (Last three lines). \textbf{Boldface} marks the best performing scenarios on each dataset.}
   \resizebox{!}{1.7cm}{    
   \begin{tabular}{lrr}
    \toprule
          & \multicolumn{1}{l}{PSNR} & \multicolumn{1}{l}{SSIM\%} \\
    \midrule
   w/o Modified latent diffusion network & 27.1074 & 90.1268 \\
    w/o Structural guidance & 27.7542 & 91.4865 \\
    w/o Auto-weight adaptation & 26.3896 & 89.7129 \\
    w/o Similar cooperative filtering & 27.9753 & 92.1584 \\
    \midrule
    \midrule
    T1 (BRATS) & 26.6355 & 91.7438\\
    T1+T1ce (BRATS) & 27.3089 & 92.9772\\
    T1+T1ce+Flair (BRATS) & \textbf{28.3126} &\textbf{93.7041}\\
    T1 (IXI) & 32.1640 & 96.0253\\
    T1+PD (IXI) & \textbf{32.8721} & \textbf{96.5932}\\
    \bottomrule
    \end{tabular}%
    }
  \label{ablation}%
\end{table}%

To test the generalizability of CoLa-Diff under the condition of varied inputs, we performed the task of generating T2 on two datasets with progressively increasing input modalities (Table \ref{ablation} bottom). Our results show that our model performance increases with more input modalities: SSIM has a maximum uplift value of 1.9603, PSNR rises from 26.6355 dB to 28.3126 dB in BRATS; from 32.164 dB to 32.8721 dB in IXI. The results could further illustrate the ability of CoLa-Diff to exploit multi-modal information.

\section{Conclusion} 
This paper presents CoLa-Diff, a DM-based multi-modal MRI synthesis model with a bespoke design of network backbone, similar cooperative filtering, structural guidance and auto-weight adaptation. Our experiments support that CoLa-Diff achieves state-of-the-art performance in multi-modal MRI synthesis tasks. Therefore, CoLa-Diff could serve as a useful tool for generating MRI to reduce the burden of MRI scanning and benefit patients and healthcare providers.

\bibliographystyle{splncs04}
\bibliography{references}
\end{document}